\documentclass[nofootinbib,twocolumn,showpacs,preprintnumbers,pre,aps,superscriptaddress]{revtex4-1}

\usepackage[dvipdfmx]{graphicx}
\usepackage{bm}
\usepackage{amsmath}
\usepackage{amssymb}
\usepackage{color}

\begin{document}

\title{Simulations of Odd Microswimmers}

\author{Akira Kobayashi}
\affiliation{
Department of Chemistry, Graduate School of Science,
Tokyo Metropolitan University, Tokyo 192-0397, Japan}

\author{Kento Yasuda}
\affiliation{
Research Institute for Mathematical Sciences, 
Kyoto University, Kyoto 606-8502, Japan}

\author{Li-Shing Lin}
\affiliation{
Department of Chemistry, Graduate School of Science,
Tokyo Metropolitan University, Tokyo 192-0397, Japan}

\author{Isamu Sou}
\affiliation{
Department of Chemistry, Graduate School of Science,
Tokyo Metropolitan University, Tokyo 192-0397, Japan}

\author{Yuto Hosaka}
\affiliation{
Max Planck Institute for Dynamics and Self-Organization (MPI DS), 
Am Fassberg 17, 37077 G\"{o}ttingen, Germany}

\author{Shigeyuki Komura}\email{komura@wiucas.ac.cn}
\affiliation{
Wenzhou Institute, University of Chinese Academy of Sciences, 
Wenzhou, Zhejiang 325001, China} 
\affiliation{
Oujiang Laboratory, Wenzhou, Zhejiang 325000, China}
\affiliation{
Department of Chemistry, Graduate School of Science,
Tokyo Metropolitan University, Tokyo 192-0397, Japan}

\begin{abstract}
We perform numerical simulations of odd microswimmers consisting of three spheres 
and two odd springs. 
To describe the hydrodynamic interaction, both the Oseen-type and the 
Rotne-Prager-Yamakawa (RPY)-type mobilities are used. 
For the Oseen-type mobility, the simulation results quantitatively reproduce the asymptotic 
expression of the average velocity.
For the RPY-type mobility, on the other hand, the average velocity is smaller than that of 
the Oseen-type mobility and the deviation is more pronounced for larger spheres.
We also perform simulations of microswimmers having different sphere sizes and show 
that the average velocity becomes smaller than that of the equal size case. 
The size of the middle sphere plays an important role in determining the average velocity.
\end{abstract}

\maketitle

\section{Introduction}

Recently, Scheibner \textit{et al.} introduced the concept of odd elasticity that is useful to characterize 
nonequilibrium active systems~\cite{Scheibner20,Fruchart22}. 
Odd elasticity arises from antisymmetric (odd) components of the elastic modulus tensor
that violates the energy conservation law and thus can exist only in active materials~\cite{Braverman21}
or biological systems~\cite{Tan22}.
Unlike passive materials, a finite amount of work can be extracted in odd elastic systems through quasi-static 
cycle of deformations~\cite{Scheibner20,Fruchart22}.
It was also shown that antisymmetric parts of the time-correlation functions in odd Langevin systems are 
proportional to the odd elasticity~\cite{Yasuda22}.
The concept of odd elasticity can be further extended to quantify the nonreciprocality of 
active micromachines (such as enzymes or motor proteins) and 
microswimmers~\cite{Yasuda22JPSJ,Kobayashi22}.
According to Purcell's scallop theorem for microswimmers~\cite{Purcell77}, nonreciprocal 
body motion is required for locomotion in a Newtonian fluid.
Within the Onsager's variational principle~\cite{DoiSoftMatterPhysics}, it was shown that odd elastic 
moduli are proportional to the nonequilibrium force~\cite{LYIHK22}.

Moreover, we have proposed a model of a thermally driven microswimmer in which three spheres 
are connected by two springs having odd elasticity~\cite{Yasuda21}. 
It was theoretically shown that the presence of odd elasticity leads to a directional locomotion 
of the stochastic microswimmer.
We have analytically obtained the average velocity under the assumption that the sphere size 
and the spring extensions are small enough compared to the natural length of the spring.
As we show later again, the average velocity is proportional to the odd elastic constant whose 
sign determines the swimming direction.

In this paper, we report the results of numerical simulations of odd microswimmers to check the 
validity of our analytical prediction~\cite{Yasuda21}.  
For comparison, we use both the Oseen-type and the Rotne-Prager-Yamakawa (RPY)-type 
hydrodynamic mobilities in our simulations~\cite{Rotne69,Yamakawa70}.
To numerically integrate the multiplicative Langevin equations, we also employ the previously 
suggested formulation that assures the equilibrium distribution~\cite{Lau07,Kuroiwa14}. 
Although our previous work considered only the case when the sphere size is identical~\cite{Yasuda21}, 
we also perform the simulations for odd microswimmers having different sphere sizes.

\section{Model}

\begin{figure}[t]
\begin{center}
\includegraphics[scale=0.35]{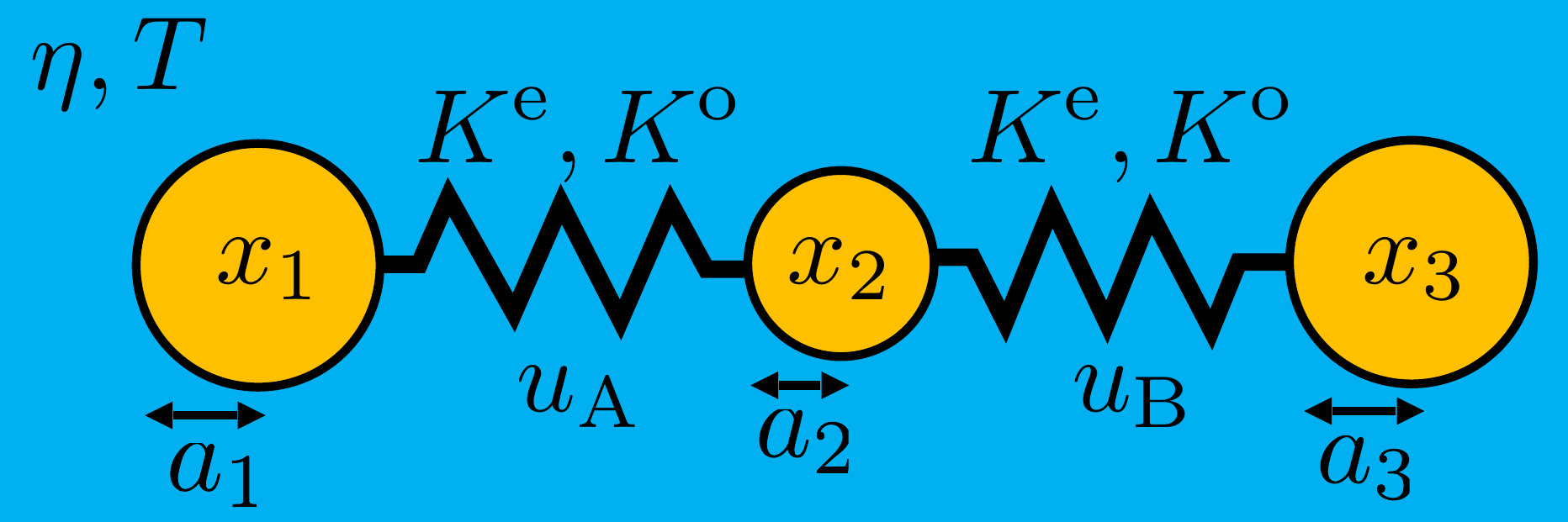}
\end{center}
\caption{
(Color online) Odd microswimmer in a fluid with viscosity $\eta$ and temperature $T$. 
Three spheres of radius $a_i$ ($i=1,2,3$) are connected by two springs with natural length $\ell$.
Each spring has both even elastic constant $K^\mathrm e$ and odd elastic constant 
$K^\mathrm o$. 
}
\label{Fig:model}
\end{figure}

Let us first review the model of an odd microswimmer~\cite{Yasuda21}.
As depicted in Fig.~\ref{Fig:model}, we consider a three-sphere microswimmer in which the positions of 
the three spheres of radius $a_i$ are given by $x_i$ ($i=1, 2, 3$) in a one-dimensional coordinate 
system~\cite{Golestanian08,GolestanianCargo08}. 
These three spheres are connected by two springs that exhibit both even and odd 
elasticities~\cite{Scheibner20,Fruchart22}.
We denote the two spring extensions by 
$u_\mathrm{A}=x_2-x_1-\ell$ and $u_\mathrm{B}=x_3-x_2-\ell$, where $\ell$ is the natural length.
Then the forces $F_\mathrm A$ and $F_\mathrm B$ conjugate to $u_\mathrm A$ and 
$u_\mathrm B$, respectively, are given by $F_\alpha=-K_{\alpha\beta}u_\beta$ 
($\alpha, \beta = \mathrm A, \mathrm B$).
For an odd spring, the elastic constant $K_{\alpha\beta}$ is given 
by~\cite{Yasuda21,Yasuda22JPSJ,Yasuda22,Kobayashi22} 
\begin{align}
K_{\alpha\beta}=K^\mathrm{e}\delta_{\alpha\beta}+K^\mathrm{o}\epsilon_{\alpha\beta}, 
\end{align}
where $K^\mathrm{e}$ and $K^\mathrm{o}$ are even and odd elastic constants
in the two-dimensional configuration space, $\delta_{\alpha\beta}$ is the Kronecker delta, 
and $\epsilon_{\alpha\beta}$ is the 2D Levi-Civita tensor.
The forces $f_i$ acting on each sphere are given by 
$f_1=-F_\mathrm A$, $f_2=F_\mathrm A-F_\mathrm B$, and $f_3=F_\mathrm B$.
We note that these forces satisfy the force-free condition, i.e., $f_1+f_2+f_3=0$.

The above odd microswimmer is immersed in a fluid of shear viscosity $\eta$ and 
temperature $T$. 
Then the Langevin equation of each sphere is given by 
\begin{align}
\dot x_i=M_{ij} f_j+\xi_i,
\end{align}
where $\dot x_i=dx_i/dt$ and $M_{ij}$ are the hydrodynamic mobility coefficients.
In the previous work, we used the Oseen-type mobility~\cite{Yasuda21}
\begin{align}
M_{ij}^{\rm O}=\begin{cases}
    1/(6\pi\eta a_i) & (i=j) \\
    1/(4\pi\eta \, |x_i-x_j|) & (i \ne j)
\end{cases}.
\label{OMobility}
\end{align}
The Gaussian white-noise sources $\xi_i$ have zero mean, 
$\langle\xi_i (t) \rangle=0$, and their correlations satisfy the 
fluctuation-dissipation theorem
$\langle\xi_i(t)\xi_j(t')\rangle=2k_\mathrm B T M_{ij}^{\rm O}\delta(t-t')$, 
where $k_\mathrm B$ is the Boltzmann constant. 
The total velocity of the microswimmer is given by $V=(\dot x_1+\dot x_2+\dot x_3)/3$.

When all the spheres have the same radius $a$, the average velocity was obtained in the limit of 
$u_\mathrm A, u_\mathrm B \ll \ell$ and $a \ll \ell$ as~\cite{Yasuda21}
\begin{align}
\langle V \rangle \approx \frac{7k_\mathrm BT\lambda}{48\pi \eta \ell^2},
\label{velocity}
\end{align}
where $\lambda=K^\mathrm o/K^\mathrm e$. 
As discussed in Ref.~1, $\langle V \rangle$ in Eq.~(\ref{velocity}) is the product of the 
geometrical factor, the explored area in the configuration space,  and the speed of the 
rotational probability flux.

In the current simulation study, we also use the RPY-type mobility that takes into account 
the next-higher order term in the far-field approximation~\cite{Rotne69,Yamakawa70}:
\begin{align}
M_{ij}^{\rm RPY}=\begin{cases}
1/(6\pi\eta a_i) \\
(i=j) \\
\dfrac{1}{4\pi\eta |x_i-x_j|} \left( 1- \dfrac{a_i^2+a_j^2}{3\vert x_i-x_j \vert^2} \right) \\
(i \ne j~{\rm and}~\vert x_i-x_j \vert \ge a_i+a_j) & \\
\dfrac{1}{6\pi\eta \tilde{a}_{ij}} \left( 1- \dfrac{3\vert x_i-x_j \vert}{16 \tilde{a}_{ij}} \right) \\
(i \ne j~{\rm and}~\vert x_i-x_j \vert < a_i+a_j) &
\end{cases}.
\label{RPYMobility}
\end{align}
The last expression is used when the spheres overlap each other, and we employ 
$\tilde{a}_{ij}=(a_i+a_j)/2$ for the effective Stokes radius~\cite{Carrasco99}.
Although this is not the unique form of the effective Stokes radius~\cite{Carrasco99,Wajnryb13,Zuk14},
the above choice is sufficient for our simulation because the motion of the microswimmer is 
restricted to a one-dimensional space.

\section{Simulation Method}

The above coupled stochastic differential equations are multiplicative because the noise
amplitudes depend on the particle positions.  
To solve such Langevin equations numerically, we use the It\^{o} interpretation ${x_i}^*=x_i(t)$ and 
integrate the quantities~\cite{Lau07,Kuroiwa14}
\begin{align}
dx_i  & = M_{ij} (x^*) f_j (x^*) dt+k_{\rm B}T \frac{\partial M_{ij} (x^*)}{\partial x_j}dt
\nonumber \\
& +\sqrt{2k_{\rm{B}}T\Lambda_{k \ell}(x^*)} Q_{ik}(x^*) (Q^{-1})_{\ell j}(x^*) dW_j,
\label{eq:motion_sim}
\end{align}
where $dW_j \sim \sqrt{dt}$ is the increment of a Wiener process and 
$\Lambda_{ij}=(Q^{-1})_{i k} M_{k \ell} Q_{\ell j}$ is the 
diagonalized matrix [$(Q^{-1})_{ij}$ being the inverse matrix of $Q_{ij}$].
The second term (proportional to $k_{\rm B}T$) on the right-hand side of 
Eq.~(\ref{eq:motion_sim}) guarantees the equilibrium distribution and is also required for nonequilibrium situations~\cite{Lau07,Kuroiwa14}.

\begin{figure}[t]
\begin{center}
\includegraphics[scale=0.5]{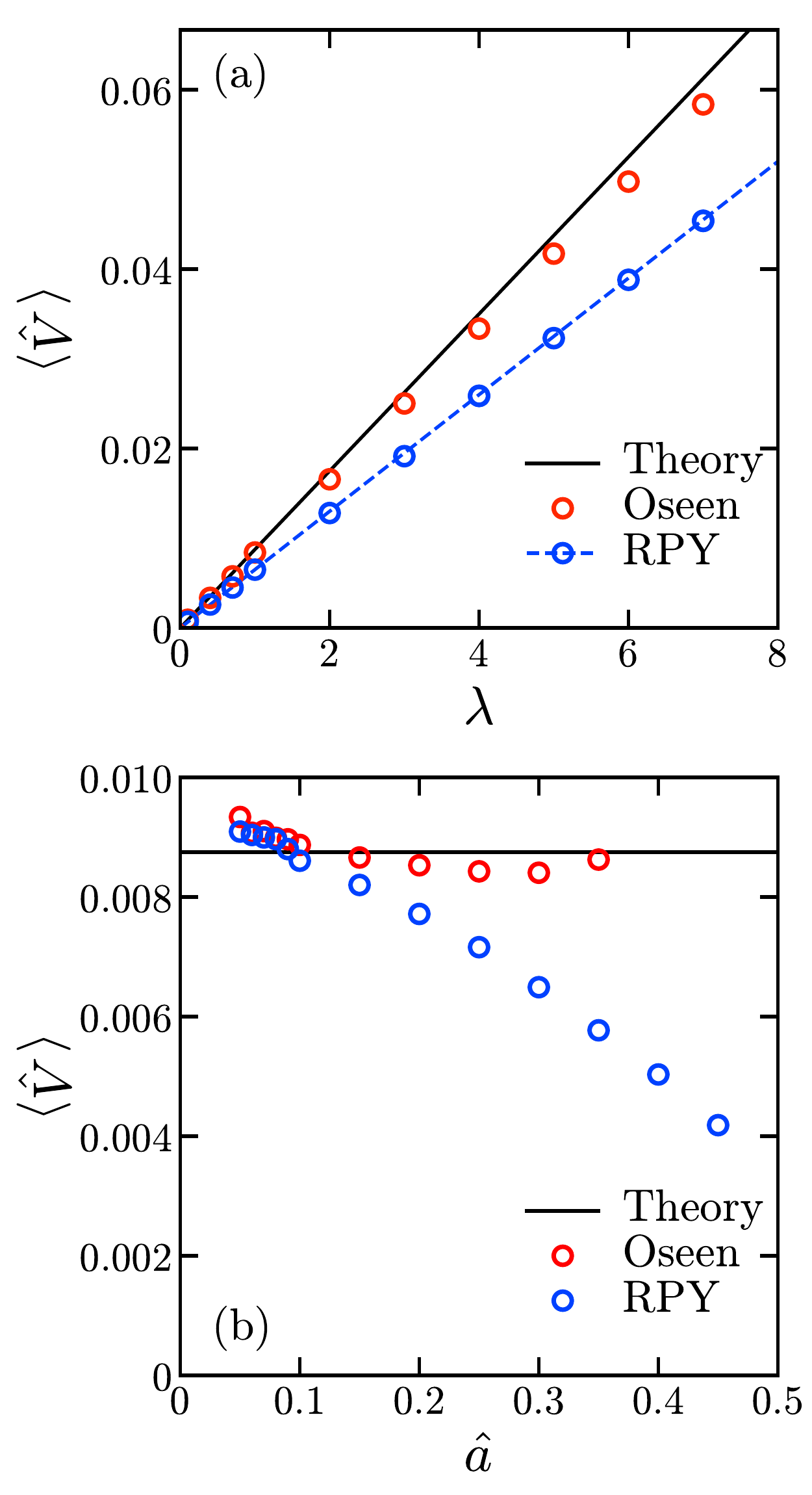}
\end{center}
\caption{
(Color online)
(a) Plot of the the dimensionless average velocity 
$\langle \hat {V} \rangle = \langle V \rangle \tau/\ell$ ($\tau=6\pi\eta \ell/K^\mathrm e$)
as a function of $\lambda=K^\mathrm{o}/K^\mathrm{e}$ when $\hat{a}=0.3$.
(b) Plot of $\langle \hat {V} \rangle$ as a function $\hat{a}=a/\ell$ when $\lambda=1$.
In (a) and (b), the dimensionless temperature is $\hat{T}=k_{\rm B}T/(K^{\rm e} \ell^2 )=0.01$,
and the black lines are the asymptotic expression in Eq.~(\ref{velocity}).
The red and blue circles are the simulation results obtained by using the Oseen-type 
[see Eq.~(\ref{OMobility})] and the Rotne-Prager-Yamakawa-type [see Eq.~(\ref{RPYMobility})] mobilities, 
respectively.
The blue dashed line in (a) is the linear fitting to the result of RPY mobility.
For the Oseen-type mobility in (b), the ratio of the terminated simulation runs are
0\% for $\hat{a} < 0.3$,
0.1\% for $\hat{a} = 0.3$,
36\% for $\hat{a} = 0.35$,
and 
100\% for $\hat{a} \ge 0.4$ (not plotted).
}
\label{Fig:equal}
\end{figure}

It should be noted that both of the mobility matrices $M_{ij}^{\rm O}$ and $M_{ij}^{\rm RPY}$ are 
not positive definite.
Hence, we terminate the simulation when the eigenvalue of $\Lambda_{k\ell}$ becomes negative 
because we need to take its square root in Eq.~(\ref{eq:motion_sim}).
In the actual simulations, such a situation occurred only for $M_{ij}^{\rm O}$ when $a_i$ is large 
and the distances between the spheres $|x_i-x_j|$ become small.
(More details are explained in the caption of Fig.~\ref{Fig:equal}.)
In the case of $M_{ij}^{\rm RPY}$, on the other hand, we did not encounter such a situation and 
all the simulation runs were completed without any termination.

We scale all the lengths by $\ell$ and use the spring relaxation time 
$\tau=6\pi\eta \ell/K^\mathrm e$ to make the time dimensionless, i.e., $\hat{t}=t/\tau$.
The dimensionless time increment is chosen as $d\hat{t}=0.01$ for $\lambda \le 1$
and $d\hat{t}=0.001$ for $\lambda > 1$.
The simulations were performed for two different dimensionless temperatures,
i.e., $\hat{T}=k_{\rm B}T/(K^{\rm e} \ell^2 )=0.01$ and $0.001$.
Each run consists of $10^7$ updates of the sphere positions.
To calculate the average velocity $\langle V \rangle$, we have taken the average over 
$2\times 10^3$ and $4\times 10^3$ independent runs for equal and different sphere size 
cases, respectively.

\section{Simulation Results}

\subsection{Equal Sphere Size}

We first discuss when the sphere size is identical and given by $\hat{a}=a/\ell$.
In Fig.~\ref{Fig:equal}(a), we plot the dimensionless average velocity 
$\langle \hat{V} \rangle = \langle V \rangle \tau/\ell$ as a function of 
$\lambda=K^\mathrm o/K^\mathrm e$ when $\hat{a}=0.3$ and 
$\hat{T}=0.01$. 
The black straight line is the scaled asymptotic expression in Eq.~(\ref{velocity}),
$\langle \hat{V} \rangle =7 \hat{T} \lambda/8$, whereas 
the red and blue circles are the simulation results obtained by using the 
Oseen-type ($M_{ij}^{\rm O}$) and the RPY-type ($M_{ij}^{\rm RPY}$) mobilities, respectively.
For the Oseen-type mobility, the simulation result agrees well with the asymptotic expression 
except for larger $\lambda$ values which give slightly smaller $\langle \hat{V} \rangle$.
This result confirms the validity of Eq.~(\ref{velocity}) for the above parameters.
For the RPY-type mobility, on the other hand, the simulation result is systematically 
smaller than Eq.~(\ref{velocity}) although the linear dependence on $\lambda$ is still maintained.
This is indicated by the fitted blue dashed line that has a smaller slope.
The reason for smaller $\langle V \rangle$ is that the inter-sphere interaction for $M_{ij}^{\rm RPY}$
is generally weaker than that for $M_{ij}^{\rm O}$.  
Notice that, in the analytical derivation of Eq.~(\ref{velocity}), we have used the condition $\hat{a} \ll 1$ 
while $\lambda$ does not necessarily have to be small.

In Fig.~\ref{Fig:equal}(b), we plot the average velocity $\langle \hat{V} \rangle$ as 
a function of the sphere size $\hat{a}$ when $\lambda=1$ and $\hat{T} =0.01$.
According to Eq.~(\ref{velocity}), $\langle V \rangle$
does not depend on the sphere size $a$ as shown by the black line.
For the Oseen-type mobility (red circles), the simulation result is relatively in good agreement 
with the asymptotic expression up to $\hat{a}\le 0.35$.  
When $\hat{a} > 0.4$, however, the mobility matrix is no longer positive definite and 
most of the simulation runs were terminated.  
(This is why there is no red data plotted for $\hat{a} \ge 0.4$ for the Oseen-type mobility.
The ratio of terminated simulation runs for $\hat{a} \le 0.35$ is written in the caption of 
Fig.~\ref{Fig:equal}.)
For the RPY-type mobility (blue circles), the simulation result coincides with the 
asymptotic expression up to $\hat{a} \le 0.15$.
For $\hat{a} \ge 0.2$ the simulation data systematically deviates from the asymptotic 
value and the deviation is more pronounced for larger spheres.

When we reduce the temperature to $\hat{T} =0.001$, the average 
velocity discussed above simply decreases by ten times. 
This is consistent with the fact that $\langle \hat{V} \rangle$ is proportional to 
$\hat{T}$ in Eq.~(\ref{velocity}).
Since the $\lambda$- and $\hat{a}$-dependencies are essentially the same, we do not
show the results for $\hat{T} =0.001$.

\subsection{Different Sphere Size}

\begin{table}[t]
\caption{
Dimensionless average velocity $\langle \hat V \rangle$ when the sphere sizes are different
but satisfy the condition $(\hat{a}_1+\hat{a}_2+\hat{a}_3)/3=0.2$.
The combinations of the dimensionless sphere radii are represented by $\hat{a}_1$--$\hat{a}_2$--$\hat{a}_3$.
We use the RPY-type mobility for all the cases and the other parameters are 
$\lambda=1$ and $\hat{T}=0.01$.
In case (A), we further impose the condition $\hat{a}_1=\hat{a}_3$ (fore-aft symmetric microswimmers), 
whereas all the spheres have different sphere size in case (B) (fore-aft asymmetric microswimmers).
The range of the numerical error for $\langle \hat V \rangle$ is $\pm 0.07 \times 10^{-3}$.
}
\label{tabvelocity}
\begin{center}
\begin{tabular}{lcc|cc}
\hline
 & $\hat{a}_1$--$\hat{a}_2$--$\hat{a}_3$ & $\langle \hat V \rangle \times 10^3$ &
 $\hat{a}_1$--$\hat{a}_2$--$\hat{a}_3$ & $\langle \hat{V} \rangle \times 10^3 $ \\
\hline \hline
(A) & 0.275--0.05--0.275 &  2.24  &       0.175--0.25--0.175 & 7.42 \\
      &  0.25--0.1--0.25      &  4.73   &      0.15--0.3--0.15       &  6.14 \\
     & 0.225--0.15--0.225 &  6.72  &       0.125--0.35--0.125    &  4.36 \\
     & 0.2--0.2--0.2           &  7.68  &        0.1--0.4--0.1          &  2.62 \\
\hline
(B)  & 0.2--0.1--0.3 & 4.52  &    0.3--0.1--0.2 & 4.69  \\
      & 0.1--0.2--0.3 &  5.88  &    0.3--0.2--0.1 & 5.97 \\ 
      & 0.1--0.3--0.2 &  5.38  &    0.2--0.3--0.1 & 5.47 \\
\hline
\end{tabular}
\end{center}
\end{table}

Next, we discuss the cases when the sphere sizes are different.
This is currently possible only by performing simulations because there is no corresponding 
analytical prediction.
In order to make fair comparisons, we investigate several cases for which the average 
sphere size is always fixed to $(\hat{a}_1+\hat{a}_2+\hat{a}_3)/3=0.2$.
We use the RPY-type mobility for all the cases and the parameters are set to
$\lambda=1$ and $\hat{T} =0.01$ as before.
The simulation results of the average velocity $\langle \hat{V} \rangle$ are summarized 
in Table~\ref{tabvelocity}.
For the case (A), we further impose the condition $\hat{a}_1=\hat{a}_3$ and vary the middle 
sphere size $\hat{a}_2$ (fore-aft symmetric microswimmers).
For the case (B), on the other hand, all the spheres have different sphere size (fore-aft 
asymmetric microswimmers).

\begin{figure}[t]
\begin{center}
\includegraphics[scale=0.5]{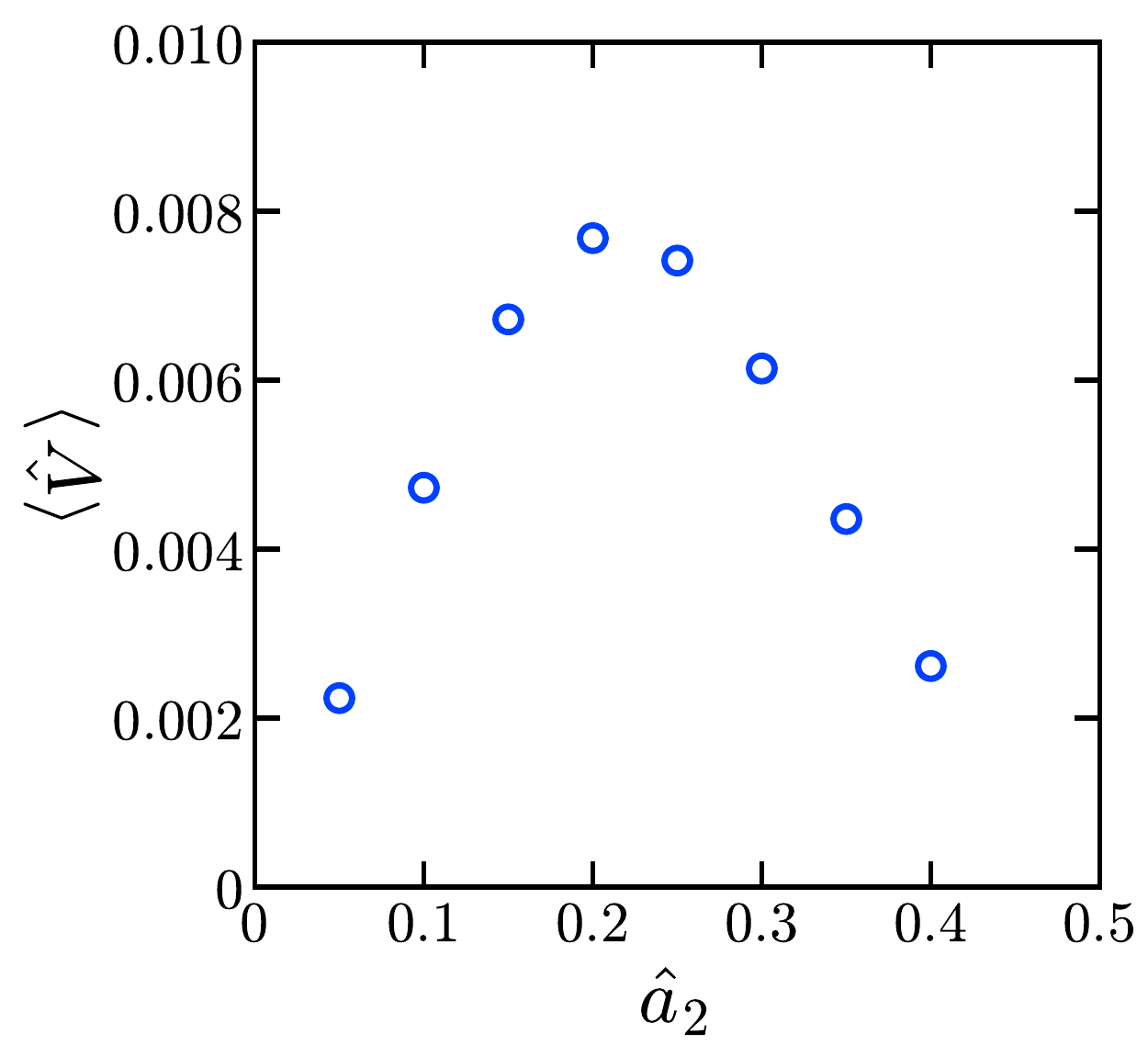}
\end{center}
\caption{
(Color online)
Plot of the the dimensionless average velocity $\langle \hat {V} \rangle$ as a function of the 
middle sphere size $\hat{a}_2$ for the case (A) in Table~\ref{tabvelocity}.
For comparison, we impose the conditions $(\hat{a}_1+\hat{a}_2+\hat{a}_3)/3=0.2$ and 
$\hat{a}_1=\hat{a}_3$ (fore-aft symmetric microswimmers).}
\label{Fig:different}
\end{figure}

In Fig.~\ref{Fig:different}, we plot  $\langle \hat {V} \rangle$ as a function of the middle sphere size 
$\hat{a}_2$ for the case (A) in Table~\ref{tabvelocity}.
Within this comparison, we clearly see that $\langle \hat {V} \rangle$ takes the largest value when 
all the spheres have the same size, i.e., $\langle \hat {V} \rangle \approx 7.68 \times 10^{-3}$ when 
0.2--0.2--0.2. 
Hence, for the fore-aft symmetric microswimmers, the size of the middle sphere essentially 
determines their average velocity.

For the fore-aft asymmetric microswimmers in case (B), on the other hand, we first note that 
$\langle \hat {V} \rangle$ is always smaller than that of the equal size case, i.e., 0.2--0.2--0.2.
It is interesting to note that $\langle \hat {V} \rangle$ almost coincides (within the error bars)
between the two asymmetric microswimmers such as 0.1--0.2--0.3 and 0.3--0.2--0.1. 
(However, such a quantitative coincidence may not hold between 0.2--0.1--0.3 and 0.3--0.1--0.2.)
Even for these fore-aft asymmetric microswimmers, the size of the middle sphere $\hat{a}_2$ 
plays an important role and the behavior of $\langle \hat {V} \rangle$ is similar to Fig.~\ref{Fig:different}.
The average velocity takes the largest and the smallest values for $\hat{a}_2=0.2$ and $0.1$, 
respectively.

The decrease of $\langle \hat {V} \rangle$ for the different sphere size cases can be physically 
understood in the following way. 
In general, the change in the average velocity can be attributed to both active and passive origins.
Within the comparison in Fig.~\ref{Fig:different}, the overall hydrodynamic friction is similar because 
the average sphere size has been fixed, leading to the similar passive contribution.  
The fact that $\langle \hat {V} \rangle$ dramatically reduces both for 
$\hat{a}_2 \gg \hat{a}_1=\hat{a}_3$ (effectively one sphere) and $\hat{a}_2 \ll \hat{a}_1=\hat{a}_3$ 
(effectively two spheres) is that the internal actuation due to odd elasticity does not operate efficiently 
for these cases.
Hence the drop of $\langle \hat {V} \rangle$ in Fig.~\ref{Fig:different} can be mainly explained 
by the lack of internal active drive in odd microswimmers having different sphere size.

\section{Summary and Discussion}

To conclude, we have performed the numerical simulations of the previously proposed 
odd microswimmers by using both the Oseen-type and the RPY-type mobilities. 
For the Oseen-type mobility, the simulation results quantitatively reproduce the 
asymptotic expression of the average velocity.
For the RPY-type mobility, on the other hand, the average velocity is smaller than that of the 
Oseen-type mobility due to the weaker hydrodynamic interactions.
Performing simulations for microswimmers having different sphere sizes, we showed 
that the average velocity becomes smaller than that of the equal size case. 
Moreover, the size of the middle sphere plays an important role in determining the average 
velocity for both fore-aft symmetric and antisymmetric microswimmers.

Here, we give a rough estimate of the parameter $\lambda=K^\mathrm o/K^\mathrm e$ that 
is the most important quantity in our model.
We consider the case when the concept of odd elasticity is applied to the structural changes of 
enzymes and motor proteins even though they are usually not microswimmers~\cite{Yasuda22,Yasuda22JPSJ,Kobayashi22}.
According to the experiments on a kinesin molecule~\cite{Ariga18}, the even elasticity can be roughly 
estimated as $K^\mathrm e \approx 10^{-4}$\,J/m$^2$.
On the other hand, the active force due to kinesin is estimated to be $f \approx 10^{-11}$\,N~\cite{Ariga18}.
By roughly estimating the natural length to be $\ell \approx 10^{-8}$~m, 
the odd elastic constant can be estimated as $K^\mathrm o\sim f/\ell \approx 10^{-3}$\,J/m$^2$.
Then the ratio between the odd and even elastic constants is typically 
$\lambda = K^\mathrm o/K^\mathrm e \approx 10$.
This is consistent with the plotted range of $\lambda$ Fig.~\ref{Fig:equal}(a).
(Notice again that $\lambda$ in Eq.~(\ref{velocity}) does not necessarily have to be small.)

Even though odd microswimmers are not yet realized experimentally, the concept of odd elasticity
has been employed in odd active robots~\cite{Ishimoto22,Brandenbourger}. 
We believe that the results of the current simulations will be useful in designing odd microswimmers
in the future.

\begin{acknowledgments}
We thank K.\ Ishimoto for useful discussions.
K.Y.\ acknowledges the support by a Grant-in-Aid for JSPS Fellows (No.\ 21J00096) from the 
Japan Society for the Promotion of Science (JSPS).
S.K.\ acknowledges the support by the National Natural Science Foundation of China (Nos.\ 12274098 and 
12250710127) and the startup grant of Wenzhou Institute, University of Chinese Academy of Sciences 
(No.\ WIUCASQD2021041).
\end{acknowledgments}



\begin{thebibliography}{99}

\bibitem{Scheibner20}
C. Scheibner, A. Souslov, D. Banerjee, P. Sur\'{o}wka, W. T. M. Irvine, and V. Vitelli, 
Nat. Phys. 16, 475 (2020).

\bibitem{Fruchart22}
M. Fruchart, C. Scheibner, and V. Vitelli, 
arXiv:2207.00071. 

\bibitem{Braverman21}
L. Braverman, C. Scheibner, B. VanSaders, and V. Vitelli, 
Phys. Rev. Lett. 127, 268001 (2021).

\bibitem{Tan22}
T. H. Tan, A. Mietke, J. Li, Y. Chen, H. Higinbotham, P. J. Foster, S. Gokhale, J. Dunkel, and N. Fakhri,
Nature 607, 287 (2022).

\bibitem{Yasuda22}
K. Yasuda, K. Ishimoto, A. Kobayashi, L.-S. Lin, I. Sou, Y. Hosaka, and S. Komura,
J. Chem. Phys. 157, 095101 (2022).

\bibitem{Yasuda22JPSJ}
K. Yasuda, A. Kobayashi, L.-S. Lin, Y. Hosaka, I. Sou, and S. Komura,
J. Phys. Soc. Jpn. 91, 015001 (2022).

\bibitem{Kobayashi22}
A. Kobayashi, K. Yasuda, K. Ishimoto, L.-S. Lin, I. Sou, Y. Hosaka, and S. Komura,
arXiv:2211.16089.

\bibitem{Purcell77}
E. M. Purcell, Am. J. Phys. 45, 3 (1977). 

\bibitem{DoiSoftMatterPhysics} 
M. Doi, \textit{Soft Matter Physics} 
(Oxford University Press, Oxford, 2013).

\bibitem{LYIHK22}
L.-S. Lin, K. Yasuda, K. Ishimoto, Y. Hosaka, and S. Komura,
arXiv.2209.15363.

\bibitem{Yasuda21}
K. Yasuda, Y. Hosaka, I. Sou, and S. Komura, 
J. Phys. Soc. Jpn. 90, 075001 (2021). 

\bibitem{Rotne69}
J. Rotne and S. Prager,
J. Chem. Phys. 50, 4831 (1969).

\bibitem{Yamakawa70}
H. Yamakawa,
J. Chem. Phys. 53, 436 (1970).

\bibitem{Lau07}
A. W. C. Lau and T. C. Lubensky,
Phys. Rev. E 76, 011123 (2007).

\bibitem{Kuroiwa14}
T. Kuroiwa and K. Miyazaki,
J. Phys. A: Math. Theor. 47, 012001 (2014).

\bibitem{Golestanian08}
R. Golestanian and A. Ajdari, 
Phys. Rev. E 77, 036308 (2008).

\bibitem{GolestanianCargo08}
R. Golestanian,
Eur. Phys. J. E 25, 1 (2008).

\bibitem{Carrasco99}
B. Carrasco, J. G. de la Torre, and P. Zipper,
Eur. Biophys. J. 28, 510 (1999).

\bibitem{Wajnryb13}
E. Wajnryb, K. A. Mizerski, P. J. Zuk, and P. Szymczak,
J. Fluid Mech. 731, R3 (2013).

\bibitem{Zuk14}
P. J. Zuk, E. Wajnryb, K. A. Mizerski, and P. Szymczak,
J. Fluid Mech. 741, R5 (2014).

\bibitem{Ariga18}
T. Ariga, M. Tomishige, and D. Mizuno, 
Phys. Rev. Lett. 121, 218101 (2018).

\bibitem{Ishimoto22} 
K. Ishimoto, C. Moreau, and K. Yasuda, 
Phys. Rev. E 105, 064603 (2022). 

\bibitem{Brandenbourger}
M. Brandenbourger, C. Scheibner, J. Veenstra, V. Vitelli, and C. Coulais, 
arXiv:2108.08837.


\end{thebibliography}
\end{document}